\documentclass{article}
\usepackage{spconf,amsmath,graphicx}

\usepackage{enumitem}
\setlist{nosep, leftmargin=14pt}

\usepackage{mwe} 

\usepackage[numbers]{natbib}

\usepackage{enumitem}
\usepackage{graphicx}
\usepackage{array}
\usepackage{hyperref}
\usepackage{colortbl}
\usepackage{amssymb}
\usepackage{booktabs}
\usepackage{amsmath}
\usepackage{float}
\usepackage{verbatim}
\usepackage{booktabs}
\usepackage{multirow}
\usepackage{makecell}
\usepackage{multicol}
\usepackage{tcolorbox}
\usepackage{color}
\usepackage{url}
\usepackage{caption}
\setlist{nosep, leftmargin=14pt}
\usepackage{pifont}
\newcommand{\cmark}{\ding{51}}%
\newcommand{\xmark}{\ding{55}}%

\usepackage{authblk}

\title{EG-SpikeFormer: Eye-Gaze Guided Transformer on Spiking Neural Networks for Medical Image Analysis}
\name{Yi Pan$^{1,*}$, Hanqi Jiang$^{1,*}$, Junhao Chen$^{1}$, Yiwei Li$^{1}$, Huaqin Zhao$^{1}$, Yifan Zhou$^{1}$, Peng Shu$^{1}$ \\ Zihao Wu$^{1}$, Zhengliang Liu$^{1}$, Dajiang Zhu$^{2}$, Xiang Li$^{3}$, Yohannes Abate$^{4}$, Tianming Liu$^{1,\dagger}$
\thanks{$^*$Equal Contribution.}
\thanks{$\dagger$Corresponding Author.}
}

\address{$^1$School of Computing, University of Georgia, Athens, GA, USA\\$^2$Department of Computer Science and Engineering, University of Texas at Arlington, TX, USA\\$^3$Department of Radiology, Massachusetts General Hospital and Harvard Medical School, MA, USA\\
$^4$Department of Physics and Astronomy, University of Georgia, Athens, GA, USA}

\begin{document}

\maketitle
\begin{abstract}
Neuromorphic computing has emerged as a promising energy-efficient alternative to traditional artificial intelligence, predominantly utilizing spiking neural networks (SNNs) implemented on neuromorphic hardware. Significant advancements have been made in SNN-based convolutional neural networks (CNNs) and Transformer architectures. However, neuromorphic computing for the medical imaging domain remains underexplored. In this study, we introduce EG-SpikeFormer, an SNN architecture tailored for clinical tasks that incorporates eye-gaze data to guide the model's attention to the diagnostically relevant regions in medical images. Our developed approach effectively addresses shortcut learning issues commonly observed in conventional models, especially in scenarios with limited clinical data and high demands for model reliability, generalizability, and transparency. Our EG-SpikeFormer not only demonstrates superior energy efficiency and performance in medical image prediction tasks but also enhances clinical relevance through multi-modal information alignment. By incorporating eye-gaze data, the model improves interpretability and generalization, opening new directions for applying neuromorphic computing in healthcare.
\end{abstract}
%

\section{Introduction}
Spiking Neural Networks (SNNs) have garnered significant attention for their bio-inspired properties and low power consumption~\cite{tavanaei2019deep,yamazaki2022spiking}. In computer vision, SNNs offer enhanced energy efficiency and interpretability compared to traditional Convolutional Neural Networks (CNNs) and Transformer architectures. However, their adoption is often hindered by lower accuracy. To overcome this limitation, researchers have combined SNNs with Artificial Neural Networks (ANNs) to leverage the strengths of both model types. Notably, integrations of SNNs with traditional CNNs~\cite{fang2021deep,zheng2021going} and advanced Transformer models~\cite{zhouspikformer,yaospike} have been explored. In many Transformer-based approaches, spike neurons replace standard neurons within the architecture, which can impede the full realization of SNNs' low-power advantages. Recent studies have proposed spike-driven transformers employing Spike-Driven Self-Attention, reducing energy consumption by utilizing linear computations at token and channel levels~\cite{zhou2023spikformer,yao2024spike}.

Despite these advancements, the application of SNNs in medical image processing remains limited. Some exceptions include the use of reservoir SNNs combined with salient feature extraction and time encoding for breast cancer image recognition, achieving high accuracy across multiple datasets~\cite{10.1155/2022/8369368}. Another study introduced a spiking convolutional neural network (SCNN) for photon-based imaging, converting time-of-flight data into high-resolution 3D images with superior accuracy and energy efficiency compared to conventional methods~\cite{kirkland2020imaging}. Additionally, SNNs have been applied to image segmentation using adaptive synaptic weights to effectively process noisy medical images~\cite{zheng2019image}. Nevertheless, the integration of SNNs in medical imaging is constrained by challenges in training and scalability relative to traditional deep learning techniques.

In this paper, we propose a novel gaze-guided spike-driven hybrid model EG-SpikeFormer for medical diagnosis for the first time, which integrates the low-power computation benefits of SNN with the powerful feature extraction capabilities of Transformers. The model incorporates radiologists' eye-gaze data as prior information during training, effectively guiding the model's attention. Our experiments on two public medical datasets demonstrate the model's superior performance in both energy efficiency and diagnostic accuracy. The main contributions of our work are as follows:

\begin{figure*}[htb]
\begin{center}
\includegraphics[trim=5 6 0 9.5,clip, width=0.8\textwidth]{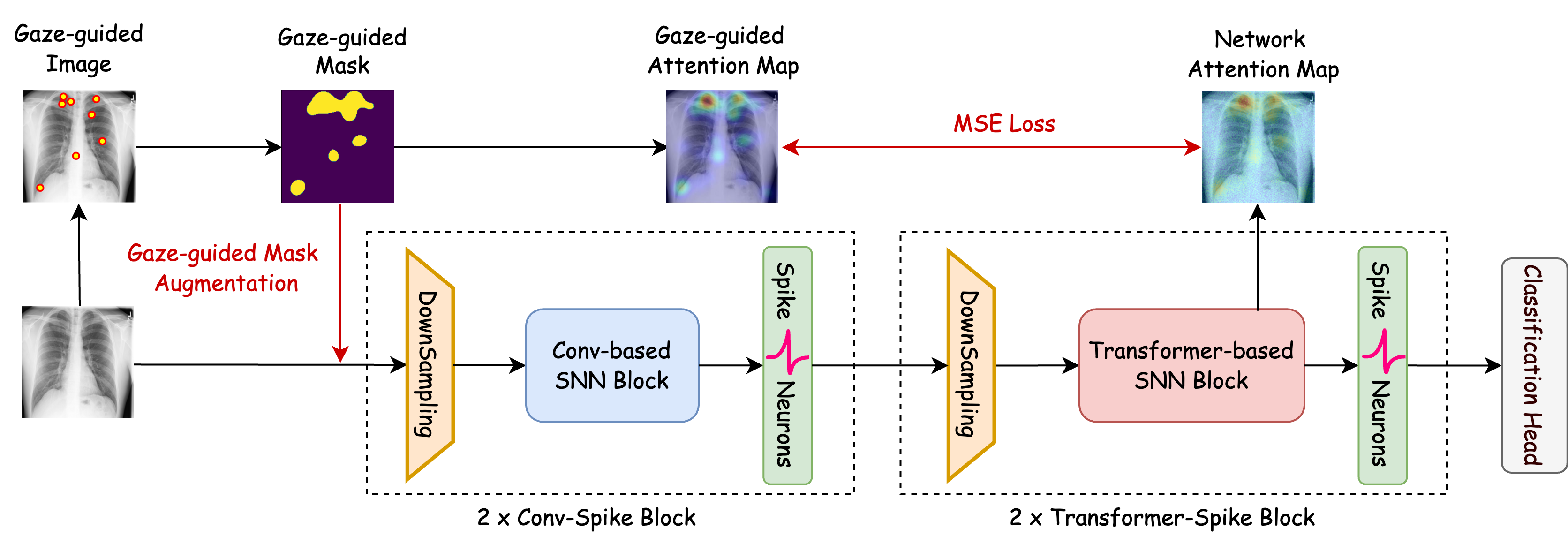}
\end{center}\vspace{-2pt}
\caption{Overall Architecture of the proposed framework.}
\label{pipeline_fig}
\end{figure*}
\begin{itemize}
\item To the best of our knowledge, this is the first attempt to introduce a hybrid SNN model combining CNN and Transformer in the field of medical diagnosis, leveraging both low power consumption and strong interpretability.
\item By incorporating radiologists' prior information during training, the model learns to focus on disease-relevant regions, reducing irrelevant information and significantly improving performance and interpretability in medical image diagnosis tasks.
\item We firstly propose a hardware-aware co-design framework that integrates neuromorphic computing and eye-gaze guidance, addressing medical challenges such as shortcut learning and data scarcity. This collaborative design not only enhances diagnostic accuracy and energy efficiency but also underscores the significance of such synergies in advancing practical healthcare applications.
\end{itemize}

\section{Method}
The overall architecture of the proposed framework is illustrated in Figure \ref{pipeline_fig}. This framework integrates eye-gaze data with a hybrid SNN to optimize both spatial and temporal feature extraction. 
\subsection{Blocks with Spike Neurons}

\begin{figure}[htb]
\begin{center}
\includegraphics[trim=10 10 170 0,clip, width=.9\columnwidth]{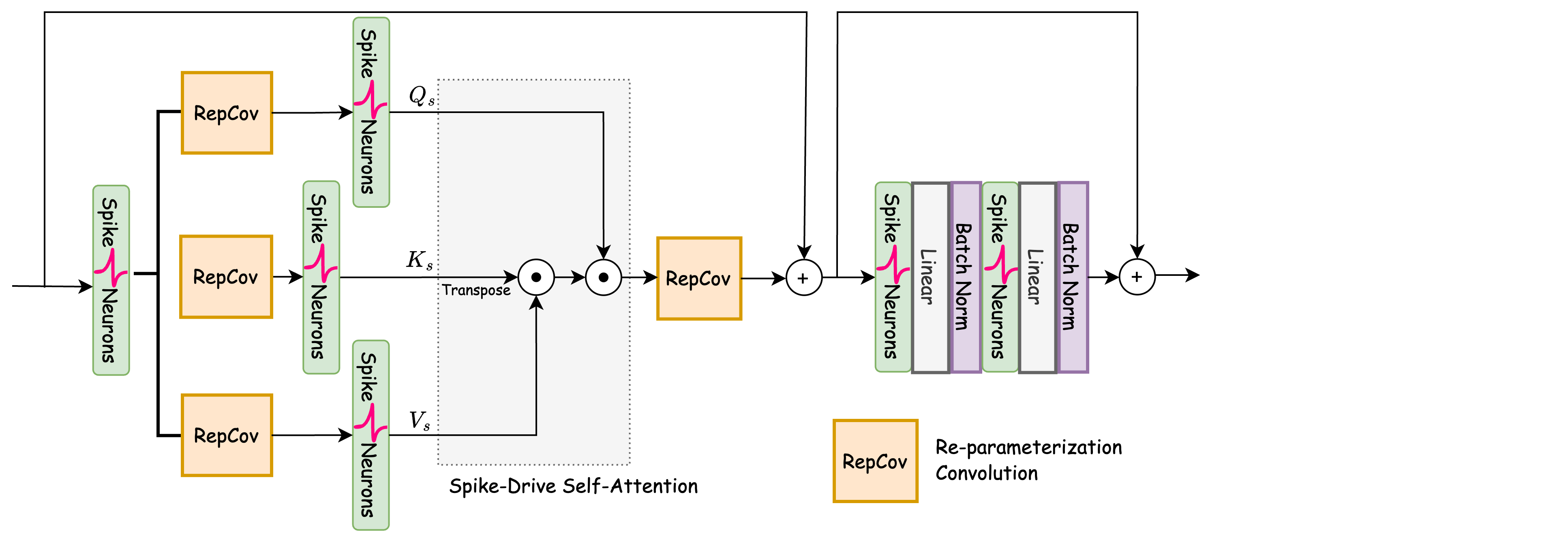}
\end{center}
\vspace{-5pt}
\caption{Architecture of the Transformer-based SNN block.}
\label{transformer_fig}
\end{figure}

Our EG-SpikeFormer utilize Leaky Integrate-and-Fire (LIF) neurons~\cite{SNN} as fundamental units in both convolution-based and Transformer-based SNN blocks. The LIF model maintains a membrane potential that accumulates incoming signals and decays over time. When this potential reaches a predefined threshold, the neuron emits a spike and resets. The dynamics of our LIF neurons are mathematically described by:
\vspace{-10pt}
\begin{align}
    V_s &= V_{t-1} + \omega x, \\
    S_t &=
    \begin{cases}
      1, & \text{if } V_s \ge V_{\text{th}} \\
      0, & \text{otherwise}
    \end{cases} \\
    V_t &= S_t V_{\text{reset}} + (1-S_t)(1 - \lambda) V_s
\end{align}
Here, \( V_{t-1} \) is the membrane potential from the previous time step, \( \omega \) is the synaptic weight, \( x \) is the input signal, \( V_s \) is the membrane potential after receiving input at time \( t \), and \( S_t \) is the binary output spike. Upon reaching or exceeding the threshold \( V_{\text{th}} \), the neuron fires (\( S_t = 1 \)) and resets its potential to \( V_{\text{reset}} \); otherwise, the potential decays with leakage factor \( \lambda \).
\begin{figure}[ht]
  \centering
  \includegraphics[trim=0 0 60 0,clip, width=.9\linewidth]{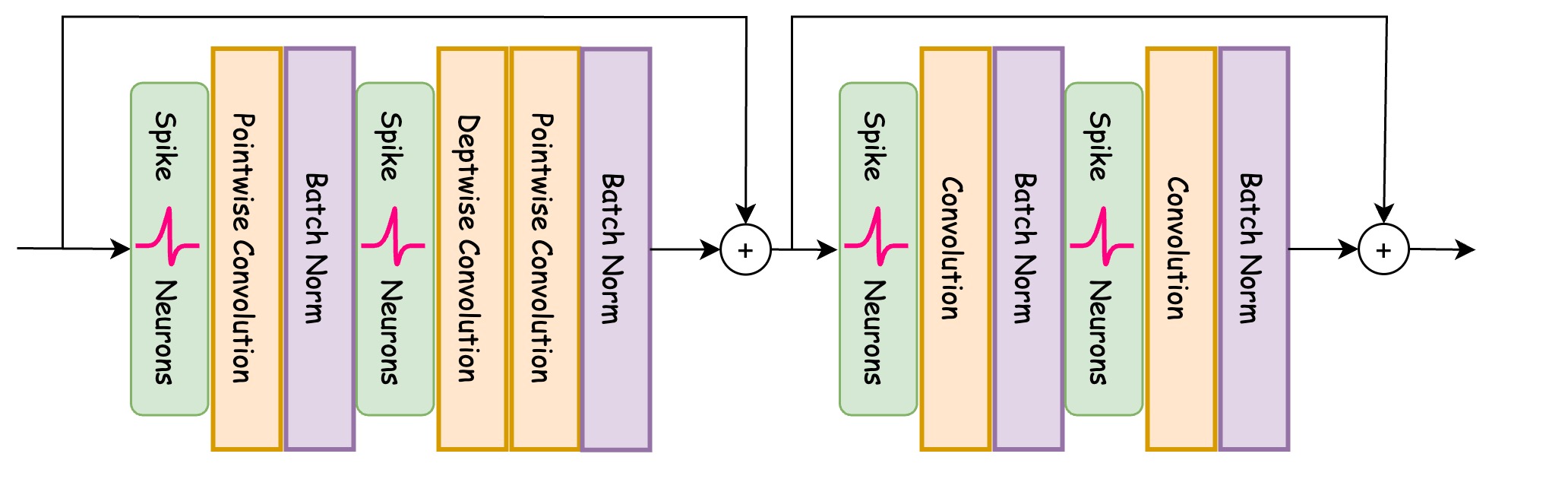}
  \caption{Architecture of the convolution-based SNN block.}
  \label{cnn_fig}
\vspace{-10pt}
\end{figure}
The convolution-based SNN block, illustrated in Figure~\ref{cnn_fig}, integrates LIF neurons with convolutional operations to enhance feature extraction and energy efficiency. It consists of spike neurons (SN), pointwise convolutions (\( \text{Conv}_p \)), depthwise convolutions (\( \text{Conv}_d \)), and batch normalization (BN) layers. Spiking activations introduce sparsity, reducing computational overhead while effectively representing spatio-temporal features. The membrane shortcut allows direct input propagation and is formulated as:
\begin{equation}
\vspace{-1pt}
x' = x + \text{BN}\left( \text{Conv}_p\left( \text{Conv}_d\left( \text{SN}(x) \right) \right) \right)
\end{equation}
The Transformer-based SNN block, depicted in Figure~\ref{transformer_fig}, integrates the attention mechanisms of traditional Transformers with the spiking neurons. In this block, spiking neurons generate spike-based queries \( Q_s \), keys \( K_s \), and values \( V_s \), which are essential components of the self-attention mechanism. These spike-based representations are obtained by processing incoming spike signals through re-parameterized convolution layers that culminate with spiking neurons. Specially, for a input $X$ to the transformer-based SNN block, the queries, keys, and values are computed as 
\begin{align}
    \{Q_s, K_s, V_s\} = \text{SN}(\text{RepCon}(X, W)) 
\end{align} 
The $W \in \{W_q, W_k, W_v\}$ are learnable weight matrices for the respective query, key, and value convolutions. Each spiking neuron SN(·) accumulates input over time and fires when its membrane potential crosses a threshold, creating sparse, event-driven representations for \( Q \), \( K \), and \( V \). The attention output is then computed using the spike attention mechanism with a shortcut. 
\vspace{-1pt}
\begin{equation}
    X' = X + \text{RepCon}((\frac{Q_s K_s^T }{\sqrt{d_k}})V_s )
\end{equation}
By combining these elements, the block effectively leverages both spatial and temporal processing capabilities, making it well-suited for dynamic or event-driven data tasks.
\vspace{-3pt}
\vspace{-15pt}
\subsection{Eye-gaze Guided Spike Vision Transformer}
Although SNNs optimize the energy efficiency of network architectures, the accuracy of SNN-based Transformers is generally lower than that of traditional Vision Transformer (ViT). SNNs inherently possess temporal and spatial characteristics, but current research predominantly focuses on natural images. In small-scale medical datasets, eye-gaze data, which contains temporal information, can serve as valuable prior knowledge for SNNs, guiding network convergence. Therefore, we designed two strategies to leverage this insight.

We augment the original images by integrating eye-gaze data to highlight key regions without losing information from other areas through the gaze mask (GM) method. Let $\mathbf{I} \in \mathbb{R}^{H \times W \times C}$ denote the original image, and let $\mathbf{M} \in \mathbb{R}^{H \times W}$ be the normalized eye-gaze mask with values between 0 and 1. The enhanced image $\mathbf{I}'$ is computed as:
\vspace{-1pt}
\begin{equation}
\mathbf{I}' = \mathbf{I} \odot (\mathbf{1} + \alpha \mathbf{M})
\end{equation}
\vspace{-1pt}
where $\odot$ denotes element-wise multiplication, $\mathbf{1}$ is an all-ones matrix of compatible dimensions, and $\alpha$ is a hyperparameter controlling the emphasis on eye-gaze regions. This operation amplifies pixel intensities in areas with higher eye-gaze values, effectively guiding the network's focus.
To align the model's attention with human gaze patterns, we introduce an attention alignment (ALH) loss. Let $\mathbf{A}_t \in \mathbb{R}^{N \times N}$ be the model's attention map from the last layer of the final Transformer block, and $\mathbf{A}_g \in \mathbb{R}^{N \times N}$ be the attention map derived from eye-gaze data, where $N$ is the number of tokens. The attention alignment loss $\mathcal{L}_{\text{align}}$ is defined as:
\vspace{-2pt}
\begin{equation}
\mathcal{L}_{\text{align}} = \frac{1}{N^2} \sum_{i=1}^{N} \sum_{j=1}^{N} \left( \mathbf{A}_t^{(i,j)} - \mathbf{A}_g^{(i,j)} \right)^2
\end{equation}
This loss encourages the Transformer module to focus on diagnostically relevant regions, improving convergence speed and performance. The total loss function combines the standard classification loss $\mathcal{L}_{\text{cls}}$ with the attention alignment loss:
\begin{equation}
\mathcal{L}_{\text{total}} = \mathcal{L}_{\text{cls}} + \lambda \mathcal{L}_{\text{align}}
\end{equation}
where $\lambda$ is a weighting factor balancing the two loss components.
\vspace{-5pt}
\section{Experimental Results and Discussion}
\vspace{-5pt}
\subsection{Datasets}
To evaluate the proposed model, we utilized two public clinical datasets: INbreast~\cite{moreira2012inbreast} and SIIM-ACR~\cite{saab2021observational,siim-acr-pneumothorax-segmentation}. The INbreast dataset comprises 410 full-field digital mammography images from low-dose X-ray breast examinations, classified into normal (302 cases), benign (37 cases), and malignant (71 cases) based on BI-RADS assessments~\cite{liberman2002breast}. Eye movement data were collected from diagnoses made by a radiologist with 10 years of experience. The SIIM-ACR dataset includes 1,170 images, with 268 cases of pneumothorax, and eye gaze data were obtained from three experienced radiologists~\cite{saab2021observational,siim-acr-pneumothorax-segmentation}. For the INbreast dataset, patients were randomly split into 80\% for training and 20\% for testing. To balance and diversify the training set, random cropping and contrast-related augmentations were applied, resulting in 482 normal, 512 benign mass, and 472 malignant mass samples. During testing, images from the remaining 20\% were processed using a sliding window with a size of 1,024 and a stride of 512. In the SIIM-ACR dataset, all images were 1,024×1,024 pixels and were directly input into the model without additional cropping.
\vspace{-5pt}
\subsection{Results and Analysis}
\begin{table*}[ht]
\centering
\resizebox{\textwidth}{!}{ 
\begin{tabular}{lccc|cccc|cccc}
\toprule
\multirow{2}{*}{\textbf{Method}} & \multirow{2}{*}{\textbf{Spike}} & \multirow{2}{*}{\textbf{Power (mJ)}} & \multirow{2}{*}{\textbf{Param (M)}} & \multicolumn{4}{c}{\textbf{INbreast \cite{moreira2012inbreast}}} & \multicolumn{4}{c}{\textbf{SIIM-ACR \cite{siim-acr-pneumothorax-segmentation}}}  \\
\cmidrule(lr){5-8} \cmidrule(lr){9-12}
& & & & \textbf{Acc.}~$\uparrow$ & \textbf{AUC}~$\uparrow$ & \textbf{F1}~$\uparrow$ & \textbf{SSIM}~$\uparrow$ & \textbf{Acc.}~$\uparrow$ & \textbf{AUC}~$\uparrow$ & \textbf{F1}~$\uparrow$ & \textbf{SSIM}~$\uparrow$ \\
\midrule
ResNet-50  \cite{he2016deep}      & \xmark & 19.0    & 26 & 89.19 & 86.62 & 80.51 & 0.276 & 84.65 & 71.97 & 83.83 & 0.153 \\
ResNet-101   \cite{he2016deep}    & \xmark & 36.1   & 45  & 90.54 & 87.84 & 88.14 & 0.302 & 84.80 & 75.55 & 84.75 & 0.158 \\
EfficientNet  \cite{tan2019efficientnet}   & \xmark & 56.6    & 119  & 91.46 & 86.62 & 88.86 & 0.352 & 82.66 & 74.37 & 79.56 & 0.197 \\
Swin-T    \cite{liu2021swin}       & \xmark & 13.8    & 28   & 91.80 & 88.10 & 90.27 & 0.227 & 84.62 & 74.82 & 84.67 & 0.159 \\
ViT     \cite{vaswani2017attention}         & \xmark & 81.0 & 86   & 92.07 & 87.13 & 86.96 & 0.395 & 84.00 & 70.76 & 83.03 & 0.205 \\
MS-Res-SNN   \cite{yao2023attention}    & \cmark & 10.2    & 77   & 80.68     & 71.92     & 69.74     & 0.178     & 83.59     & 69.92     & 81.37     & 0.202     \\
\midrule\midrule
ResNet-50+Gaze   & \xmark & 19.0    & 26   & 90.54 & 86.93 & 86.00 & 0.208 & 84.80 & 75.00 & 83.71 & 0.153 \\
ResNet-101+Gaze  & \xmark & 36.1    & 45   & 91.88 & 88.89 & 88.19 & 0.244 & 84.80 & 72.68 & 82.89 & 0.254 \\
ViT+Gaze         & \xmark & 81.0 & 86 & 93.12 & \textcolor{blue}{92.30} & 91.83 & \textcolor{blue}{0.402} & 85.60 & \textcolor{blue}{75.30} & 85.35 & \textcolor{blue}{0.280} \\
Ours-T2          & \cmark & 26.7 & 55   & \textcolor{blue}{93.18}     & 91.76     & \textcolor{blue}{92.20}     & 0.398     & \textcolor{blue}{87.89} & 75.17 & \textcolor{blue}{85.39} & 0.273 \\
Ours-T4          & \cmark & 52.4 & 55   & \textcolor{red}{\textbf{94.32}}     & \textcolor{red}{\textbf{93.00}}     & \textcolor{red}{\textbf{95.53}}     & \textcolor{red}{\textbf{0.427}}     & \textcolor{red}{\textbf{88.28}} & \textcolor{red}{\textbf{76.20}} & \textcolor{red}{\textbf{86.33}} & \textcolor{red}{\textbf{0.285}} \\
\bottomrule
\end{tabular}
}
\caption{Performance Comparison on INbreast and SIIM-ACR Datasets. The Accuracy (Acc.), ROC AUC (AUC), and F1-score (F1) metrics are reported. \textcolor{red}{\textbf{Red}} and \textcolor{blue}{blue} denote the best and second-best results.}
\label{tab:1}
\vspace{-5pt}
\end{table*}

We extensively evaluated our proposed model on the INbreast~\cite{moreira2012inbreast} and SIIM-ACR~\cite{siim-acr-pneumothorax-segmentation} datasets, comparing it with various existing methods. To assess how well the model's attention aligns with human gaze patterns, we calculated the Structural Similarity Index (SSIM) between the model's attention maps and the gaze data. Following Chen et al.~\cite{chen2023training}, we also computed the energy consumption for each model, demonstrating that our approach maintains competitive performance while being energy-efficient. The total energy consumption is expressed as a combination of the energy required for both multiply-accumulate (MAC) and accumulate (AC) operations~\cite{power_equation}:
\vspace{-3pt}
\begin{equation}
    \text{Energy} = E_{\text{MAC}} \times \text{FLOPs}(c) + E_{\text{AC}} \times \text{SOPs}(s)
\vspace{-2pt}
\end{equation}
where \( E_{\text{MAC}} = 4.6\,\text{pJ} \) and \( E_{\text{AC}} = 0.9\,\text{pJ} \) represent the energy per MAC and AC operation implemented on the 45nm hardware~\cite{power}, respectively; FLOPs and SOPs denote the number of floating-point and spike-based operations at the CNN layers \( c \) and SNN layers \( s \).

As shown in Table \ref{tab:1}, our eye-gaze guided SNN architecture, EG-SpikeFormer achieves state-of-the-art performance across all metrics, outperforming classical ViTs, eye-gaze augmented ViTs, and CNNs. Besides lower energy consumption, our model enhances performance, demonstrating efficiency and superior capability in complex medical imaging tasks. Leveraging the spatio-temporal event-driven computation of SNNs, our architecture balances high performance with low energy consumption without compromising accuracy. It surpasses traditional deep learning architectures, highlighting the potential of SNNs in resource-constrained medical environments. Integrating eye-gaze data effectively guides the model's attention, improving accuracy and convergence speed without significantly increasing computational resources. We evaluated two training processes: a two-step and a four-step. Even with two-step training, our model matches or exceeds other state-of-the-art methods. The four-step training further enhances capabilities, indicating that additional steps can unlock more potential without substantially increasing energy consumption.
\vspace{-5pt}
\subsection{Ablation Study}
\vspace{-10pt}
\begin{table}[ht]
    \begin{center}
    \resizebox{0.48\textwidth}{!}{ 
    \begin{tabular}{c c c | c c c | c c c }
    \toprule
    \multicolumn{3}{c}{\textbf{Ablation}} & \multicolumn{3}{c}{\textbf{INbreast \cite{moreira2012inbreast}}} & \multicolumn{3}{c}{\textbf{SIIM-ACR \cite{siim-acr-pneumothorax-segmentation}}}\\
    \textbf{GM} & \textbf{ALH} & \textbf{T4} & \textbf{Acc.}~$\uparrow$ & \textbf{AUC}~$\uparrow$ & \textbf{F1}~$\uparrow$ & \textbf{Acc.}~$\uparrow$ & \textbf{AUC}~$\uparrow$ & \textbf{F1}~$\uparrow$  \\
    \midrule
    \cmark & &\cmark & 92.05 & 86.34 & 85.35 & 85.55 & 71.28 & 82.02  \\
    & \cmark &\cmark & 93.08 & 86.57 & 92.15 & 87.11 & 74.39 & 84.60  \\
   \cmark & \cmark &  & 93.18 & 91.76 & 92.20 & 87.89 & 75.17 & 85.39  \\
    \cmark & \cmark & \cmark & \textbf{94.32} & \textbf{93.00} & \textbf{95.53} & \textbf{88.28} & \textbf{76.20} & \textbf{86.33}  \\
    \bottomrule
    \end{tabular}
    }
     \end{center}
     \vspace{-10pt}
    \caption{Ablation study on the INbreast and SIIM-ACR datasets. The ablation experiments test the impact of removing GM, AM, and T4 modules. Accuracy (Acc.), AUC and F1-score (F1) are reported. \textbf{Bold} indicates the best result, and $\uparrow$ means higher values are better.}
    \label{tab:ablation_study}
    \vspace{-10pt}
\end{table}
We conducted an ablation study to assess the impact of GM, ALH, and the four-step training process (T4) on model performance. The detailed results are presented in Table~\ref{tab:ablation_study}. The model without T4 also showed lower performance metrics. The highest performance was achieved when all components were included, confirming that GM, ALH, and T4 each contribute significantly to the model's effectiveness.

\vspace{-10pt}
\section{Discussion}
\vspace{-5pt}
Neuromorphic chips, leveraging SNNs as the most accessible approach, offer transformative potential for clinical diagnostics by addressing key challenges such as energy efficiency, data scarcity, and interpretability~\cite{Aboumerhi_2023,schuman2022opportunities}. EG-SpikeFormer tackles shortcut learning and focuses on diagnostically relevant features, making it ideal for real-time medical applications in resource-constrained settings and offering a forward-thinking framework for medical neuromorphic chip design. Additionally, its eye-gaze-driven, spatio-temporal event processing optimizes neuromorphic systems for clinical tasks. By enhancing diagnostic accuracy and promoting algorithm-hardware co-design, EG-SpikeFormer facilitates the development of efficient, high-performance medical chips. Futhermore, its ability to reduce data demands and computational overhead accelerates energy-efficient and scalable chip design, advancing accessible clinical diagnostics and expanding the role of neuromorphic technologies in healthcare.
\vspace{-5pt}
\section{Conclusion}
\vspace{-5pt}
EG-SpikeFormer integrates eye-gaze data with SNN to enhance medical image analysis, utilizing the energy efficiency and spatio-temporal dynamics of SNN alongside the feature extraction strengths of CNN and Transformers. By directing attention on clinically relevant regions, the model achieves superior diagnostic accuracy, interpretability, and generalizability, and addresses the limitations of conventional models. These results highlight neuromorphic computing’s potential in healthcare, providing a high-performance, low-energy solution, particularly well-suited for medical environments with critical data constraints and accuracy requirements.
\bibliographystyle{IEEEbib}
\bibliography{refs}

\end{document}